\documentclass[a4paper,11pt]{article}
\pdfoutput=1 

\usepackage{jinstpub} 
                     
\usepackage{longtable}
\usepackage{lscape}   
\usepackage{upgreek}
\usepackage[utf8]{inputenc}

\title{Hydrogen reduction of enriched germanium dioxide and zone-refining for the LEGEND experiment}

\author[a,1]{K.-P. Gradwohl \note{Corresponding author.},}
\author[b]{O. Moras,}
\author[a]{J. Janicskó-Csáthy,}
\author[b]{S. Schönert}
\author[a]{and R.R. Sumathi}

\affiliation[a]{Leibniz-Institut für Kristallzüchtung,\\Max-Born-Str. 2, 12489 Berlin, Germany}
\affiliation[b]{Physik Department, Technische Universität München,\\James-Franck-Straße 1, 85748 Garching bei München, Germany}

\emailAdd{kevin-peter.gradwohl@ikz-berlin.de}

\abstract{The LEGEND experiment, now under construction, will operate a large array of Ge detectors for the search of neutrinoless double-beta decay of $^{76}$Ge. In this paper we report on the process development for the hydrogen reduction of germanium dioxide enriched in $^{76}$Ge as part of the effort to manufacture detectors for the LEGEND experiment. The process was optimized via a kinetic un-reacted shrinking model and tested with a batch of natural GeO$_2$. 
We completed the reduction of a batch of 23 kg isotopically enriched Ge with an average yield of 99.85\%. Subsequently, the Ge was purified to intrinsic purity by zone-refining and an overall Ge yield of 99.05\% was achieved. Using an intermediate underground storage, an average cosmogenic exposure of 156 h was accumulated. Special care was taken to avoid and recycle losses during the process.}


\begin{document}
\maketitle
\flushbottom

\section{Introduction}

The existence of the neutrinoless double-beta decay ($0\upnu\upbeta\upbeta$) is of great interest for particle physics and cosmology \cite{ackermann2013gerda,Majorana2018,agostini2019}. One of the experiments aiming to search for the $0\upnu\upbeta\upbeta$ decay is the LEGEND experiment \cite{abgrall2017legend}, which will operate high-purity germanium (HPGe) detectors isotopically enriched in $^{76}$Ge. LEGEND will be realized in stages, first LEGEND-200 with up to 200 kg target mass will be built followed by LEGEND-1000 with the final goal of operating one ton of HPGe detectors. Such an experiment would have a sensitivity for the half-life of the $0\upnu\upbeta\upbeta$ decay beyond 10$^{28}$ years.

The HPGe detectors from the $^{76}$Ge enriched germanium are all custom made, 
from starting materials provided by the collaborations. The work presented here is a continuation of the effort started with the production of the GERDA Phase II detectors \cite{enrBEGe2015} and also relies on a parallel development done for the Majorana Demonstrator experiment \cite{abgrall2018}.

Optimization of the processing of the isotopically enriched material is crucial for the success of $^{76}$Ge $0\upnu\upbeta\upbeta$ experiments. The processing affects the outcome of the experiment in many ways:

\begin{itemize}
 \item  The isotopic content has to be measured and preserved during the process. The number of moles of the $0\upnu\upbeta\upbeta$-decaying isotope and not the total detector mass is used in the calculation of the sensitivity or half-life limit. Therefore care has to be taken that the material is not diluted with natural germanium. The best way to avoid this is to set up a separate processing line.

 \item The yield of a given processing step has a significant impact on the cost of the experiment. Hence, for a fixed budget it will directly affect the total detector mass of the experiment and consequently its sensitivity. The mass yield of GERDA Phase II detectors was only 53.3\%~\cite{enrBEGe2015}. The waste is being recycled for the purpose of the LEGEND experiment. The yield after continuous reprocessing can be expressed as the product of all efficiencies raised to the power of the number of recycling steps; hence small improvements will have a significant impact on the achievable target mass.
 It is also essential to avoid irrecoverable losses and to develop methods with the possibility of recycling in mind.
 
 \item Cosmogenic activation of germanium is one of the important background sources in the detector spectrum. This can be avoided by storing the material shielded from the hadronic component of the cosmic rays. In practice this means storing the germanium under a shielding of 10~m water equivalent (w.e.) or more. Each detector production campaign requires carefully planned logistics based on a network of underground storage sites. Since underground processing, which is the optimal solution, is not practical for many reasons, minimization of processing times and careful timing is essential.
 
 \item Since the final goal is production of HPGe detectors, the impurities introduced or removed prior to crystal growth will have an impact on the final yield. Therefore introduction of contaminants affecting semiconductor properties should be avoided in every step of the processing.
\end{itemize}

In our work, we addressed all these issues in order to find the optimum solution for the processing of large quantities of germanium for the LEGEND-1000 experiment. Using existing infrastructure, a small scale production line was set up at the Leibniz-Institut für Kristallzüchtung (IKZ) Berlin, Germany. With the equipment we describe below, 33~kg of enriched GeO$_2$ was processed to electronic grade germanium for later production of HPGe detectors for LEGEND. 

The feedstock for the production of commercial HPGe detectors is widely available electronic grade germanium. The material is first purified by zone-refining followed by crystal growing by the Czochralski method. Cylindrical slices are cut from the crystals and processed into semiconductor detectors.
In contrast, the production of the enriched detectors starts with the germanium dioxide (GeO$_2$) provided by the enrichment plant. Subsequently, the oxide first has to be reduced to Ge by hydrogen. This extra step can be perceived as an opportunity to achieve better control over the purity of the germanium and eventually simplify the logistics by integrating the reduction with the purification preceding the HPGe crystal growing process.

\section{Reduction}

The hydrogen reduction of GeO$_2$ is a well known process used for industrial production of germanium \cite{hoffmann1987,melcher1988}.

\begin{align}
\textrm{GeO}_2 +  \textrm{ H}_2 &\longrightarrow   \textrm{GeO} +  \textrm{ H}_2 \textrm{O}  \label{eq:chemicalReaction1} \\
\textrm{GeO} + \textrm{ H}_2 &\longrightarrow  \textrm{Ge} +  \textrm{ H}_2 \textrm{O} \nonumber \\[-10pt]
\cline{1-2}
\textrm{GeO}_2 + 2 \textrm{ H}_2 &\longrightarrow  \textrm{Ge} + 2 \textrm{ H}_2 \textrm{O} \label{eq:chemicalReaction}
\end{align}

As an intermediate step in the reduction, GeO is formed (eq. \ref{eq:chemicalReaction1}) which sublimates at temperatures above 700~\textdegree C. Therefore, the reduction has to be completed below this temperature before melting the germanium to cast in ingots. As shown by the summary equation \ref{eq:chemicalReaction}, the reaction products are germanium and water vapor.  

\subsection{Reduction furnace}

The main features we require from a reduction furnace for the purpose of the LEGEND experiment are: a quartz (fused silica) tube instead of the standard alumina based ceramic, a closed system that facilitates the full recovery of the germanium lost during the standard process, and a process capacity of at least 1~kg of germanium per reduction. The quartz tube has several advantages over the standard alumina ceramic. Quartz is extremely resistant to thermal shock as opposed to alumina, which means one can speed up the heating and cooling procedure and consequently reduce the exposure of the enriched Ge. Furthermore, aluminum is a critical element in the HPGe purification process (see for example \cite{haller1976}) and one potential source of Al could be the reduction furnace.

\begin{figure}[htbp]
\includegraphics[width=.6\textwidth]{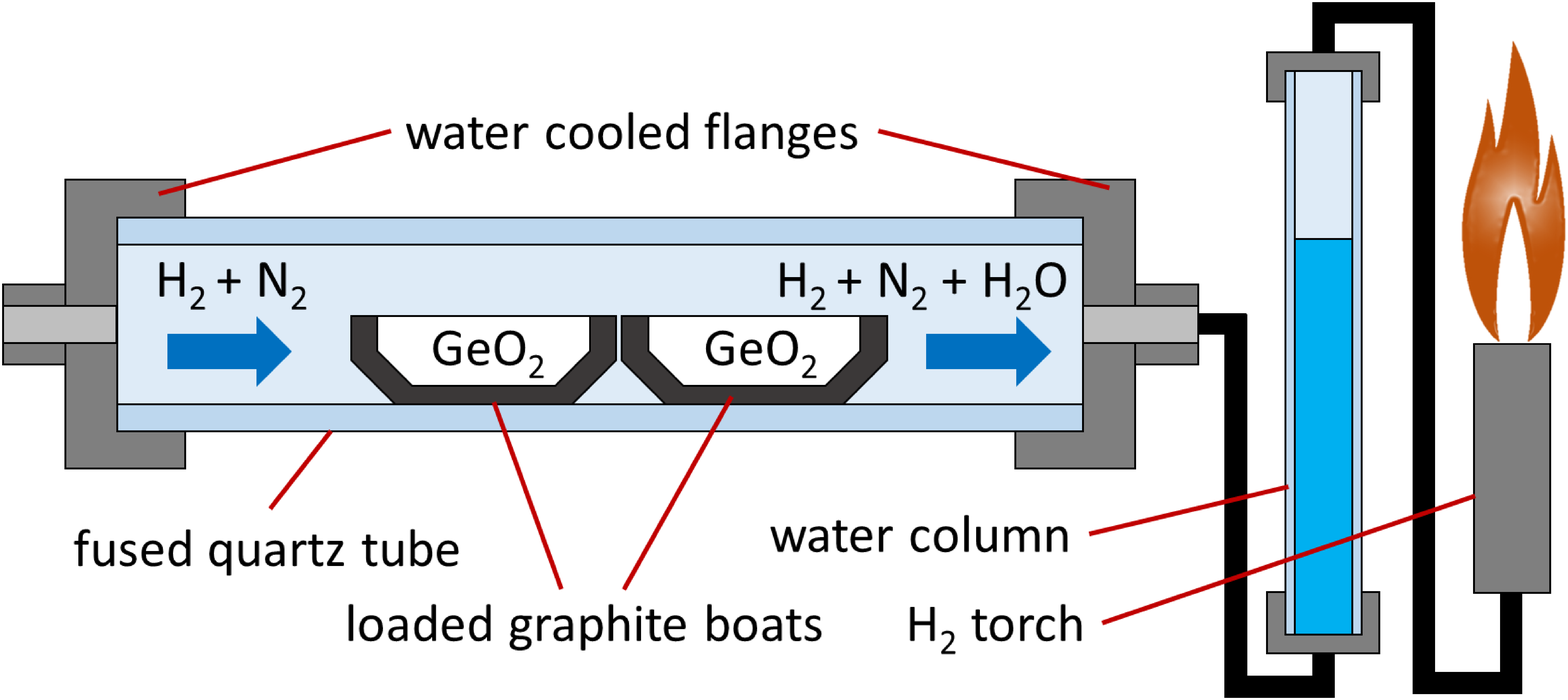}
\qquad
\includegraphics[width=.4\textwidth]{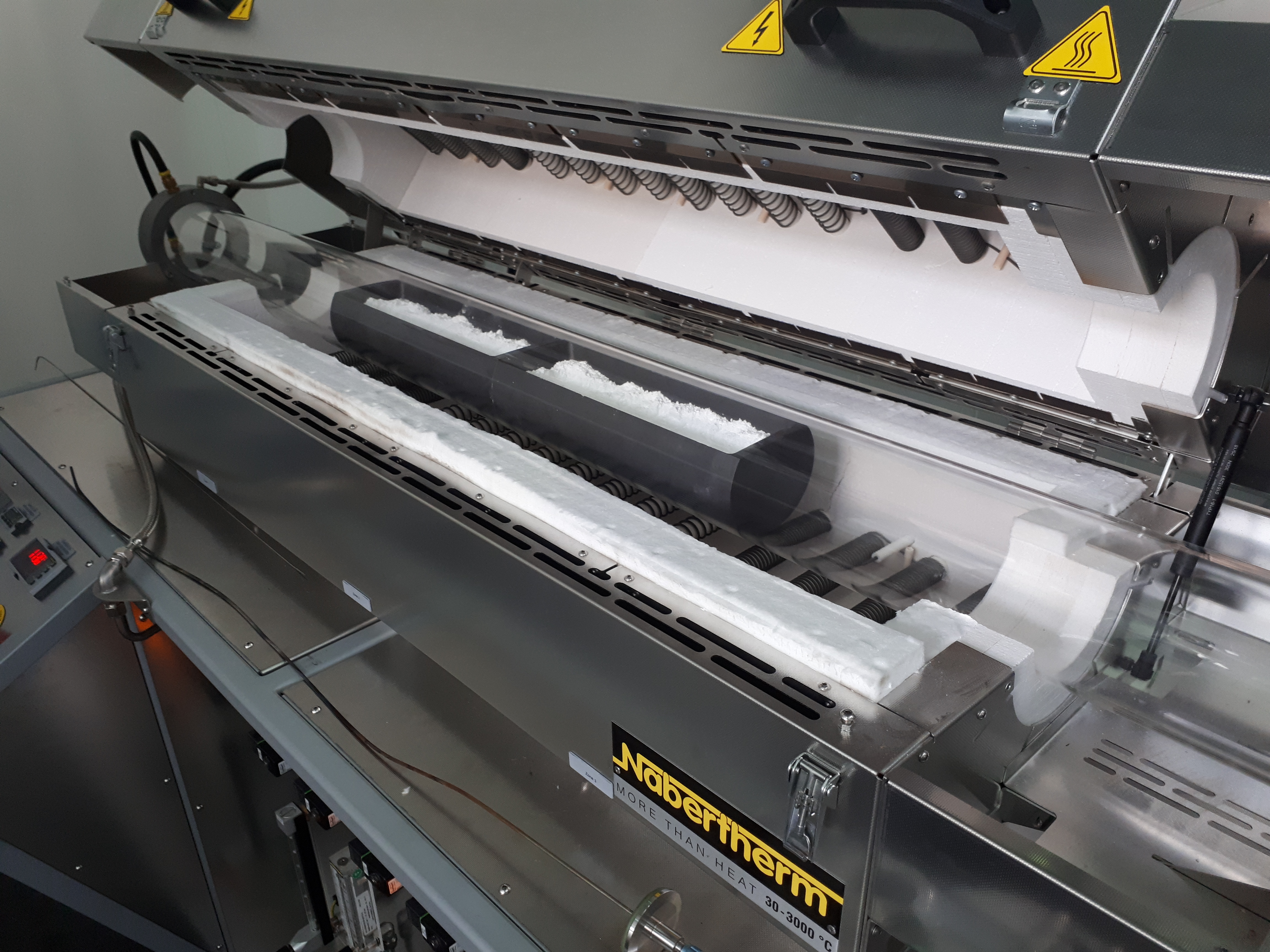}
\caption{A basic sketch (left) and the actual furnace (right) used in the GeO$_2$ hydrogen reduction process. The graphite boats are loaded with GeO$_2$ which is reduced in a mixture of H$_2$ and N$_2$. The reduction takes place at temperatures above 500~\textdegree C, measured above the graphite boats.}
\label{fig:redFurnace}
\end{figure}

The choice fell on a furnace produced by the company Nabertherm GmbH mainly because of the lower price compared to the competition. A basic schematic as well as a photograph of the reduction furnace is shown in Fig.~\ref{fig:redFurnace}. The tube furnace itself is an of-the-shelf product of the company, a split furnace type oven with a 120~mm diameter quartz tube. The furnace has three-zone resistive heating with a total heated length of 1~m and a constant temperature zone of about 500 mm.

The tube furnace was delivered mounted on the top of a bench containing the gas system and a programmable logic controller (PLC). The PLC is required for the safe operation with H$_2$ gas, as it controls the gas system and the heating power. The H$_2$ containing exhaust gas is fed to an excess gas burner (H$_2$ torch) which is also monitored by the PLC.

As an add-on option a thermocouple introduced in the tube measures the temperature of the charge directly. Along with the three other thermocouples of the three heated zones, this additional temperature sensor is used by the PLC for more precise process control. The temperatures given in this paper were always measured inside the tube.

The ends of the furnace tube are closed gas tight with water cooled metal flanges which are connected to a gas system designed for operation with H$_2$ gas. To ensure that no air can enter the tube during operation, producing an explosive mixture, the furnace operates between 30 and 120~mbar over-pressure regulated by a water column as shown in Fig.~\ref{fig:redFurnace}. This relatively simple system has the advantage that the pressure regulator also functions as a condenser for the water produced in the reduction process. Consequently, all the Ge which is transported from the boats towards the exhaust (e.g. in the form of GeO at elevated temperatures) has to desublimate within the furnace, ultimately in the water column. This Ge can then be easily recovered by evaporating the water and adding the recovered GeO$_2$ to the next reduction process.

\subsection{Process development}

Before proceeding with the reduction of the enriched GeO$_2$, the reduction process was tested and optimized with natural GeO$_2$.
Following \cite{abgrall2018} and \cite{hoffmann1987}, we first reduced one kg of GeO$_2$ at constant 650~\textdegree C for about eight hours. We experienced large quantities of water condensing in the furnace tube causing violent pressure fluctuation when water flooded the heated zone. In addition the germanium was not fully reduced in the expected eight hours. This caused further problems when we tried to heat the charge to the melting point of the Ge: GeO reduced in the gas phase was clogging the exhaust with gray Ge powder. By consequence, we realized the need for a systematic study of the reaction kinetics.  

Several fundamental investigations regarding reaction \ref{eq:chemicalReaction} have been carried out, especially concerning the thermodynamic equilibrium and the reaction kinetics \cite{dennis1923,hasegawa1972,hasegawa1972highPressure}. However, these fundamental experiments were conducted at milligram scale loads and a kilogram scale reduction process brings new challenges. In particular, for the purpose of the reduction of the enriched germanium dioxide we have to minimize the exposure to cosmic rays and hence we have to shorten the time of reduction as much as possible while still maintaining a yield above 99\%.

In principle the reduction could also be done with a constant temperature profile, but the resulting reaction rate would be very high at the beginning of the reaction (safety hazard due to pressure buildup) and very low at the end (risk of incomplete reduction). For this purpose we calculated the reaction rate of the process to optimize the temperature profile during the reduction towards a constant reaction rate (constant water production). 

The change between reducing and oxidizing conditions in equilibrium of the GeO$_2$ - H$_2$ - Ge - H$_2$O system is at around 575 to 600~\textdegree C \cite{yokokawa1957}. Regarding the reaction kinetics one can apply standard models, which are known from metal oxide hydrogen reduction processes, like the un-reacted shrinking model, as derived in \cite{dang2013}. The model is based on the assumption that the oxide particles are spherical in shape and homogeneous. The hydrogen reduces the surface of the GeO$_2$ particle and produces a product layer. For further reduction the hydrogen has to diffuse through this layer, which is gradually growing in size, causing a decrease in reaction rate at a given temperature. 

\begin{figure}[htbp]
\centering
\includegraphics[width=0.8\textwidth]{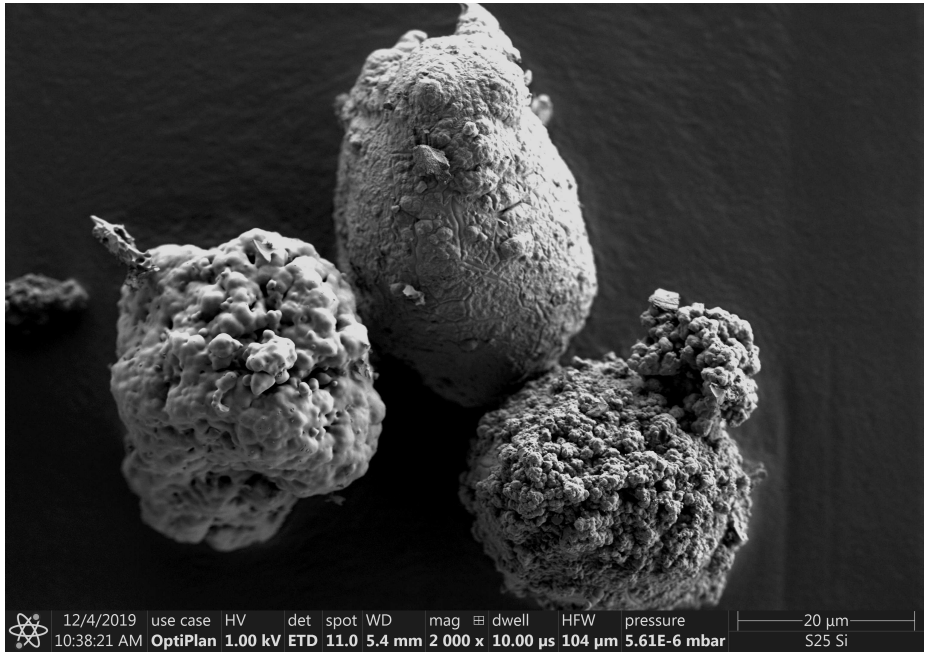}
\caption{Scanning electron microscopy image of enriched GeO$_2$ powder particles (the lower two). It can be seen that the GeO$_2$ particles are not single crystalline and that they are highly porous (high surface area). The upper particle is a Ge particle that can be found on the boat walls after reduction.}
\label{fig:rem}
\end{figure}

For amorphous germanium dioxide the hydrogen reduction is of first order reaction type, while for the hexagonal $\alpha$-quartz type GeO$_2$ it is of a different auto-catalytic type \cite{baba1956}. For a first order reaction this can be expressed with a simple differential equation in the reaction extent $R$ based on geometrical considerations, with the following analytical solution:

\begin{equation} \label{eq:reductionModel}
    R = 1 - \left(1 - C_0 \exp\left(-\frac{E_{\textrm{a}}}{kT}\right) t\right)^3,
\end{equation}

where $E_{\textrm{a}}$ is the activation energy of the chemical reaction \ref{eq:chemicalReaction}, $k$ is the Boltzmann constant, $T$ is the temperature, $t$ is the time, and $C_0$ is a temperature independent constant. $C_0$ is theoretically composed of several factors, such as the particle diameter and the hydrogen partial pressure. The powder particle diameter of the enriched GeO$_2$ was determined by light microscopy to be 42~$\upmu$m with a standard deviation of 10~$\upmu$m, based on a 200 particle statistic. It was possible to rule out that the powder is of the $\alpha$-quartz type modification by investigation with scanning electron microscopy. For this purpose the powder particles were transferred to conductive double sided adhesive carbon tabs and measured directly without conductive coating. The powder particles are of amorphous nature and show a highly porous morphology, as can be seen in Fig.~\ref{fig:rem}. The commercial natural germanium dioxide particles had a similar morphology, but were much more agglomerated and hence had a larger diameter. The upper particle of the three is a reduced and melted Ge pearl as it occurs on the boat walls after every process, showing a smooth morphology and grain boundaries.

For general determination of the reaction rate all reaction steps have to be considered. However, it could be shown by \cite{hasegawa1972,hasegawa1972highPressure} that the GeO$_2$ - Ge interface reaction is the reaction rate limiting step, and that this un-reacted shrinking model works very well to describe the reduction process. The activation energy was determined to be 77~kJ/mol. Therefore, within the reduction process described, the only fit parameter is $C_0$ which in this case contains information about the particle morphology and the hydrogen partial pressure.

\begin{figure}[htbp]
\includegraphics[width=1\textwidth,trim=60 0 60 0,clip]{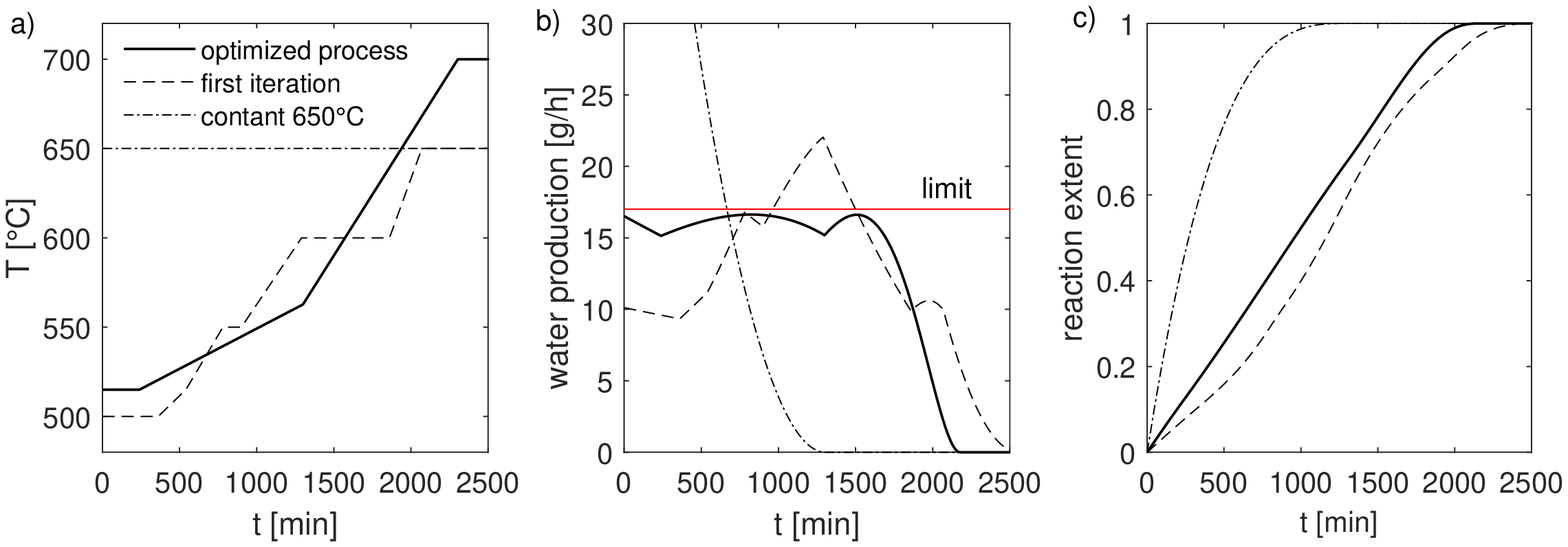}
\caption{Temperature profile of the reduction process (a), the quantity of water produced during the reduction of 1700~g GeO$_2$ (b), and the reaction extents (c) for a constant temperature profile at 650 \textdegree C, the first working but not optimized iteration, and the final optimized reduction process. The critical limit for water production is marked in (b), above which water accumulation and sudden evaporation can cause an emergency shutdown of the furnace.}
\label{fig:opti}
\end{figure}

In the first successful reduction process we slowly increased the temperature from 500~\textdegree C, holding the temperature whenever the water level became critical. In the following, this test served as the starting point of the optimization.     
A temperature profile of this first reduction can be seen in Fig.~\ref{fig:opti}a (dashed line). Based on the proposed model in Eq.~\ref{eq:reductionModel} and the assumption that the reaction is finished after the process (R = 1), it was possible to determine the value of $C_0$ to 17.4~min$^{-1}$ for the present conditions. In this model the heat capacity of the furnace and the charged boats is neglected, which is a good approximation if the heat ramps are low (as it was the case during the actual reduction process).

The calculated reaction rate expressed as water production rate and the reaction extent is depicted in Fig.~\ref{fig:opti}b and c, respectively. From the first trials it was possible to determine the critical water production rate at which water accumulation in the quartz tube is too high (17~g/h). Subsequently, the perfect temperature control profile was determined based on a two ramp model with a constant temperature incubation and ending sequence (Fig.~\ref{fig:opti}). The incubation period proposed is related to the exchange of contaminating gas adsorbed to the oxide particles with hydrogen, and the formation of metal nuclei on surface defects of the oxide particles \cite{hasegawa1972}.

At the last step of the reduction we waited at 700~\textdegree C until all the water in the tube had evaporated to ensure that all the oxide had been reduced. The final optimized reduction process (without melting the bars, heating and cooling ramps) can be finished within 36~h for 1700~g Ge without reaching the critical water accumulation limit, assuring a stable process.

GeO$_2$ is strongly hydrophilic and adsorbs water vapor from the air. It was shown by measuring adsorption isotherms that this water is present as a liquid water layer on the GeO$_2$ particles \cite{law1955}. During the numerous test runs we noticed the appearance of water condensation in the tube shortly after the heating was turned on, even without enabling the H$_2$ flow. This led us to the conclusion that the GeO$_2$ contained a significant amount of water. By drying the GeO$_2$ at 500 \textdegree C under 100~l/h N$_2$ flow for three hours, we determined the water content to be around 2\% of the weight of the natural oxide. This was taken into account when determining the process yields.

\begin{figure}
\includegraphics[width=0.9\textwidth]{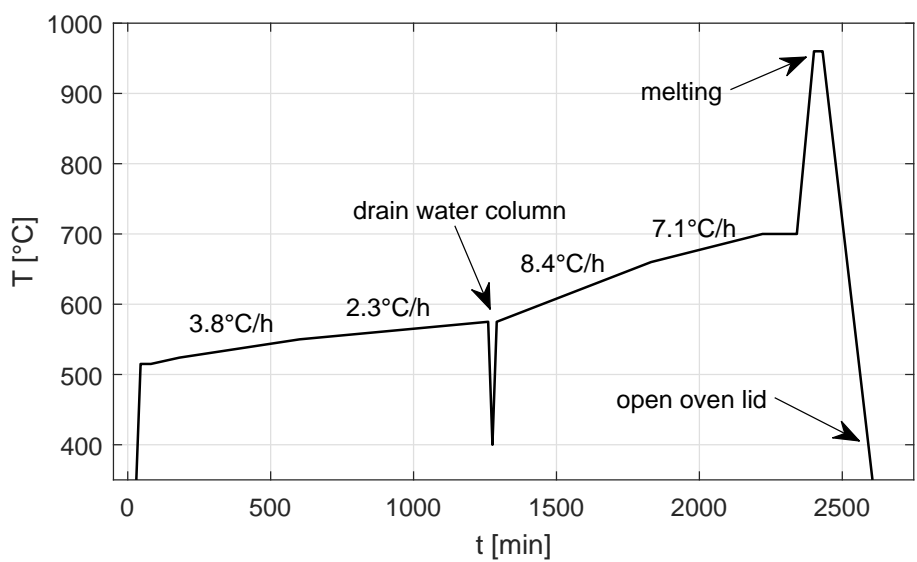}
\caption{The set temperature profile of one complete reduction process. The accumulated water in the water column is drained in the middle of the process (1260~min). At the end, the reduced Ge powder is melted to Ge bars. The complete process takes 46~h from charging the furnace with GeO$_2$ until collecting the cooled down Ge bars.}
\label{fig:TempProf}
\end{figure}

During the process development we gradually increased the amount of GeO$_2$ and were able to reduce up to 2~kg GeO$_2$ per process in two graphite boats with about 1.3~l volume.

To maintain a stable and continuous gas flow during the process the hydrogen flow was adjusted to 200~l/h. It was observed that by adding 100~l/h N$_2$ flow to the H$_2$ helped to remove water thus further stabilizing the process. 
The final process including all heating and cooling ramps and the melting procedure is shown in Fig.~\ref{fig:TempProf}. It was possible to reduce the process time to 46~h from loading the furnace with oxide to collecting the Ge bars. The water formed in the reaction collects in the water column and needs to be drained once it is full after about 21~h of process time. In the final period of the reduction (two hours at a constant temperature of 700~\textdegree C) the condensation of water on the water cooled exhaust flange stops, indicating that all the germanium dioxide has reacted to germanium (fine gray powder). The melting procedure is initialized with a steep ramp of 400~\textdegree C/h to 960~\textdegree C, where the Ge powder is melted to bars. Finally, the hydrogen flow is shut off and the tube is flushed with N$_2$. When the furnace temperature has fallen below 400~\textdegree C the oven lid is opened, greatly reducing the time needed to reach room temperature.

During test runs with natural GeO$_2$ we demonstrated that a process time under two days is possible with a process yield of 99.6\% after subtracting the water content. This result was comparable or better than that achieved by an industrial company \cite{enrBEGe2015}, therefore we decided to start the reduction of the enriched material.

\subsection{Processing of the enriched Ge}

The material used for this work had been isotopically enriched in $^{76}$Ge up to 87\% by the JSC Zelenogorsk Electrochemical Plant (ECP), Russia. The Ge we received was enriched in three different batches. Although the enriched Ge is still the same chemical substance, there were minor differences between the commercial GeO$_2$ and the batch delivered by ECP.

Using the standard atomic mass of Ge to calculate the stoichiometric ratios in GeO$_2$ will cause an error at the percent level when dealing with the isotopically enriched material. Hence, for an accurate estimation of the process yield the atomic weight was determined by inductively coupled plasma mass spectrometry (ICP-MS). The isotopic compositions of each batch are shown in Table~\ref{tab:isotopicComposition}, as determined by the Laboratori Nazionali del Gran Sasso (LNGS) chemistry laboratory. The resulting mean atomic mass of 75.74 is significantly higher than the atomic mass of natural Ge. The difference in isotopic composition between the batches is within the accuracy of the ICP-MS measurements, and hence an average atomic mass of 75.74 is used to determine the yields of the reduction processes.

\begin{table}[!ht]
\centering
\begin{tabular}{|c|c|c|c|c|c|c|c|}
\hline
\textbf{Batch} & \textbf{$^{70}$Ge [\%]}  & \textbf{$^{72}$Ge [\%]}   & \textbf{$^{73}$Ge [\%]}    & \textbf{$^{74}$Ge [\%]}     & \textbf{$^{76}$Ge [\%]}   & \textbf{A$_{\textrm{Ge}}$} \\ \hline
53/5176       & <0.01 & <0.01 & <0.01 & 13.5 $\pm$ 0.5 & 86.5 $\pm$ 1.0 & 75.73  \\ \hline
53/5177       & <0.01 & <0.01 & <0.01 & 12.6 $\pm$ 0.5 & 86.9 $\pm$ 1.0 & 75.75  \\ \hline
53/5189       & <0.01 & <0.01 & 0.019 & 13.4 $\pm$ 0.5 & 86.6 $\pm$ 1.0 & 75.73  \\ \hline
\end{tabular}
\caption{The isotopic composition of the three enriched germanium batches determined by the LNGS chemistry laboratory via ICP-MS. Furthermore, the mean atomic mass is within the measurement tolerance shown in the last column.}
\label{tab:isotopicComposition}
\end{table}

Compared to the batch of natural GeO$_2$ used for process development we noticed significant differences in the powder quality. First, the powder density of the enriched GeO$_2$ was significantly lower. Hence, by filling the boats to the maximum we could only reduce 1700~g oxide per run. We explain this with the lower particle diameter of the enriched powder.

Further we found that the water content of the powder was negligible. As with the natural GeO$_2$ we carried out a drying experiment. The weight difference before and after drying was 2.6~g with a 1422.4~g GeO$_2$ load. This difference includes the desorbed water of both graphite boats and therefore the water content in the GeO$_2$ is considered negligible. Thus, no drying procedure was applied for the subsequent processes and the incubation time at 515~\textdegree C was reduced. Consequently, all the reduction process yields are determined without the consideration of the water content of the powder.

The enriched GeO$_2$ came in $\sim$~710~g packages double sealed in polyethylene bags. The furnace was loaded with two graphite boats containing up to 850~g GeO$_2$ each. Usually, the measured weight of the oxide was about 1~g less than what was specified on the bag by ECP. The difference can be explained by the precision of the scales and the small amount of powder sticking to the inside of the bag.

Each of the 20 runs was assigned a process number based on the date of the process start. The achieved yield was >~99\% in each reduction run as can be seen in the summary in Table~\ref{tab:red}. After each reduction the furnace was reloaded with oxide and the process was restarted. 
 
\begin{table}[!ht]
\centering
\begin{tabular}{|l|l|l|l|l|}
\hline
\textbf{Process No.} & \textbf{GeO$_2$ load [g]}   & \textbf{Ge bars [g]}   & \textbf{Yield [\%]}   & \textbf{Exposure [kg$\cdot$h]}   \\ \hline
R20190930       & 1419.8 & 997.2   & 99.90  &  82.6  \\ \hline
R20191007       & 1602.2 & 1124.9  & 99.86  &  73.8  \\ \hline
R20191009       & 1699.4 & 1193.0  & 99.85  &  64.5  \\ \hline
R20191011       & 1702.7 & 1193.6  & 99.71  &  86.5  \\ \hline
R20191014       & 1703.1 & 1195.8  & 99.87  &  60.6  \\ \hline
R20191016       & 1701.4 & 1194.0  & 99.82  &  85.5  \\ \hline
R20191018       & 1700.8 & 1193.7  & 99.83  &  95.1  \\ \hline
R20191021       & 1700.8 & 1195.0  & 99.95  &  59.3  \\ \hline
R20191023       & 1701.1 & 1194.6  & 99.89  &  69.4  \\ \hline
R20191025       & 1700.6 & 1194.3  & 99.90  &  90.3  \\ \hline
R20191028       & 1700.4 & 1194.0  & 99.89  &  82.6  \\ \hline
R20191030       & 1701.3 & 1194.2  & 99.85  &  73.8  \\ \hline
R20191101       & 1701.0 & 1194.1  & 99.86  &  64.5  \\ \hline
R20191105       & 1701.1 & 1194.5  & 99.88  &  64.5  \\ \hline
R20191107       & 1700.8 & 1193.6  & 99.83  &  64.5  \\ \hline
R20191112       & 1700.7 & 1193.8  & 99.85  &  82.6  \\ \hline
R20191114       & 1701.0 & 1193.0  & 99.76  &  73.8  \\ \hline
R20191118       & 1702.3 & 1194.4  & 99.80  &  75.1  \\ \hline
R20191120       & 1423.4 & 998.7   & 99.80  &  65.3  \\ \hline
R20191122       & 1137.3 & 800.1   & 99.95  &  61.5  \\ \hline
\textbf{Total:} & \textbf{32801.2} & \textbf{23026.6}  & \textbf{99.85}  &  \textbf{1647.7}  \\ \hline
\end{tabular}
\caption{A summary of all reduction processes with the initial oxide charge, the total mass of the resulting Ge bars, the process yield, and the specific exposure of the Ge (see Section \ref{sec:exposure}).
}
\label{tab:red}
\end{table}

Since GeO$_2$ has a solubility of 4.5~g/l in water at room temperature, oxide can be recovered from the water produced in the reduction process which was collected and drained from the pressure regulating water column. Furthermore, to recover as much powder as possible, the bags were washed out in deionized water with a resistivity of 17~M$\Omega$cm. After boiling of the water 107~g oxide was recovered, which was added to the last reduction. 

After each reduction process small Ge droplets with various diameters remained on the walls of the boats (upper particle in Fig.~\ref{fig:rem}). These droplets amounted to ca. 0.5~g per run. They were collected and added to the last boat together with the recycled oxide. 

The total mass of the GeO$_2$ powder weighted in the boats was 32801.2~g. After the recovery of the Ge grains, the powder from the bags and the dissolved GeO$_2$ from the condensed water, 23026.6~g of Ge was obtained resulting in a total reduction yield of 99.85\%. Consequently, only 33.4~g of Ge was lost during the reduction process. A thin Ge layer gradually developed where the quartz tube protrudes the thermal insulation. The tube had to be etched twice to avoid damage. We expect that some of the residual 33.4~g of Ge can be regained from this solution.

\section{Zone-refining}

Zone refinement is a well documented procedure for purifying various metals \cite{Pfann:1966}. A short molten zone is moved slowly through the length of the Ge ingot, hence the name zone-refining (ZR). Since most dissolved elements in Ge have a segregation coefficient different from unity, the impurities tend to travel towards one or the other end of the bar. The tails of the ingots can be cut off and the remaining material will be a few orders of magnitude cleaner than the starting material. 

The HPGe production process requires electronic grade (6N) germanium as starting material. The purity is assessed via a resistivity measurement. The intrinsic resistivity of germanium is about 50~$\Omega$cm at room temperature and it is generally accepted that this corresponds to a chemical purity of 6N or better. The reduced material is far from that, being between 1 to 10~$\Omega$cm. 

Our self built zone refiner at IKZ consists of four tubes with one RF coil each. The ZR was conducted in pure H$_2$ atmosphere supplied from a Pd diffusion cell. The total time above ground of the enriched GeO$_2$, including the time spent for ZR, has to be minimized. Hence, the zone velocity and the number of cycles in the ZR are constrained. From the processing of the natural germanium we learned that ZR for only 24~h gives unsatisfactory results on occasion. For this reason we ran the ZR for 48~h with a zone-velocity of 5~mm/min, resulting in 16 passes. This procedure resulted in a stable yield of around 70\% of 50~$\Omega$cm Ge material. 

\begin{figure}[htbp]
\includegraphics[width=.48\textwidth,trim=0 0 0 300,clip]{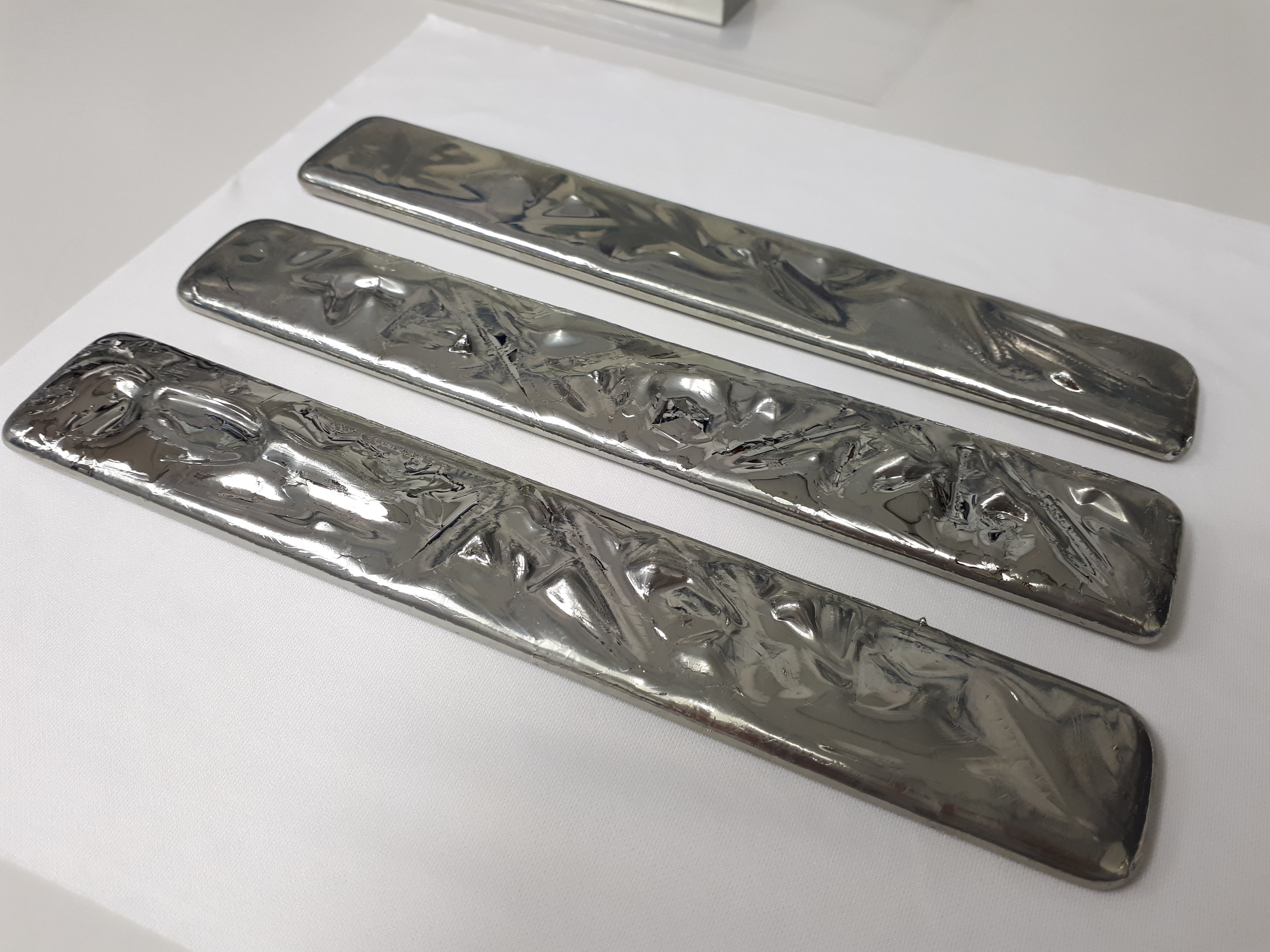}
\qquad
\includegraphics[width=.48\textwidth,trim=0 0 0 300,clip]{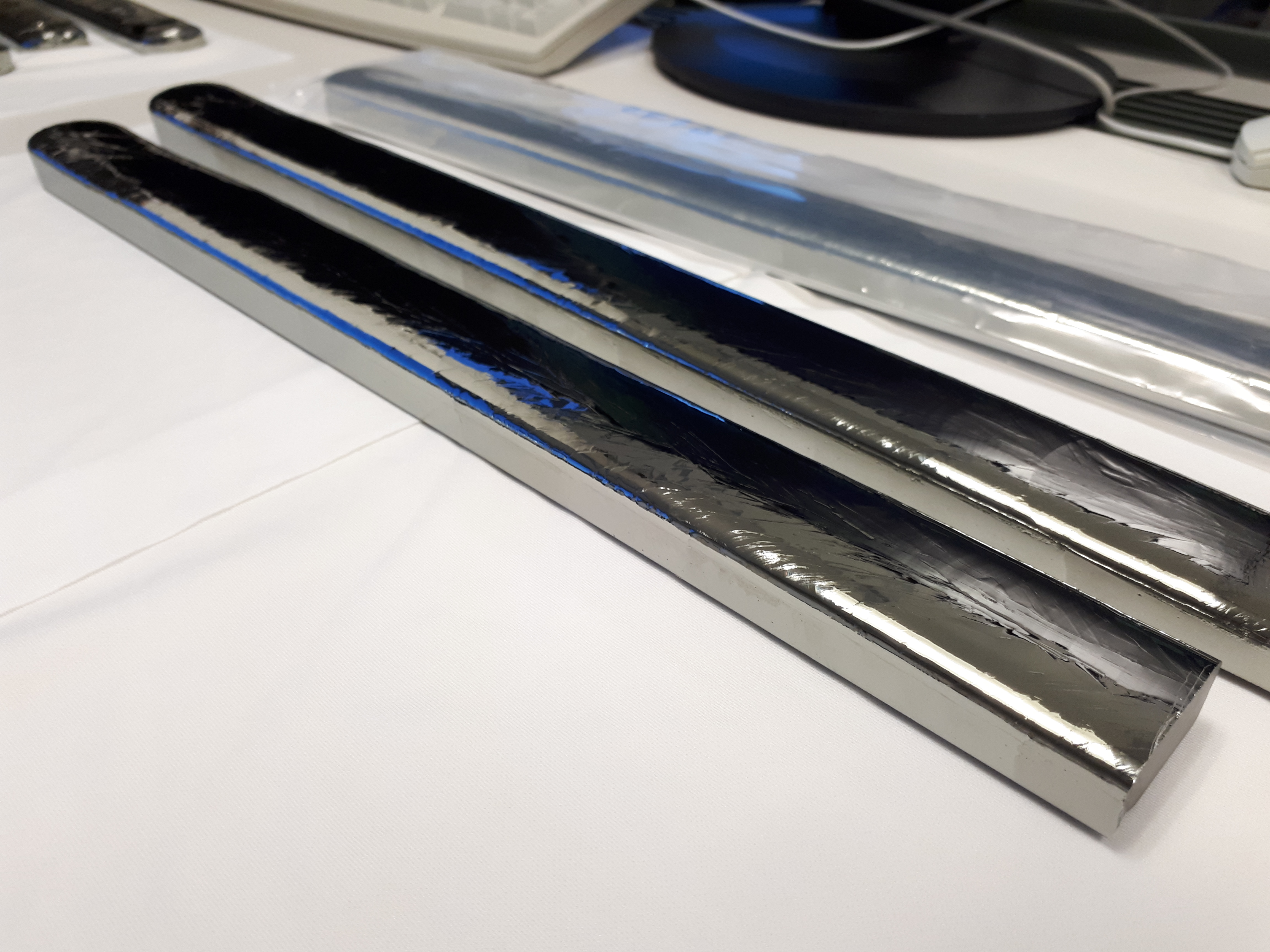}
\caption{Cleaned 600~g Ge bars from the reduction furnace (left) and finished 50~$\Omega$cm Ge bars after zone-refining and cutting (right).}
\label{fig:bars}
\end{figure}

For each ZR run two graphite boats were loaded with four Ge bars from the reduction furnace ($\sim$~2.4~kg). When a sufficient amount of freshly reduced Ge was available a ZR run was started. The standard procedure requires etching of the reduced Ge bars prior to ZR~\cite{hoffmann1987}. To avoid losses due to etching we instead washed the bars with isopropyl alcohol and 17~M$\Omega$cm deionized water. The cleaned reduction bars can be seen in Fig.~\ref{fig:bars}~(left). The high ZR yield showed that this cleaning procedure was sufficient.

The resistivity along every ZR bar was measured with a four-point measurement method. Where the resistivity dropped below 50~$\Omega$cm the bars were cut with a cut-off grinding machine with water (without lubricant) as cooling agent. This machine is dedicated to Ge cutting only, and it was cleaned thoroughly from natural Ge before cutting the enriched Ge for possible recovery of the cutting losses from the sludge.

When enough of the low resistivity tails were collected, a boat was filled with them. After ZR some of these tails even showed signs of precipitations on the last centimeter. To increase the efficiency of the recovery, the low resistivity tails were etched, as in the standard procedure in an HF (40\% by weight):HNO$_3$ (60\% by weight), (1:3) solution.

\begin{table}[!ht]
\centering
\begin{tabular}{|l|l|l|l|l|}
\hline

\multicolumn{5}{|c|}{\textbf{Bars from reduction}} \\ \hline

   & \textbf{Load mass [g]}  & \textbf{Mass 50 $\Omega$cm [g]} &  \textbf{Yield 50 $\Omega$cm [\%]} & \textbf{Exposure [kg$\cdot$h]}  \\ \hline

ZR 1 & 2193.1 & 1567.1 & 71.46 & 112.6   \\ \hline
ZR 2 & 2317.7 & 1871.5 & 80.75 & 129.4   \\ \hline
ZR 3 & 2388.4 & 1874.4 & 78.46 & 133.4   \\ \hline
ZR 4 & 2388.0 & 1780.7 & 74.57 & 127.8   \\ \hline
ZR 5 & 2388.8 & 1752.4 & 73.36 & 127.8   \\ \hline
ZR 6 & 2388.0 & 1773.4 & 74.26 & 136.3   \\ \hline
ZR 7 & 2387.6 & 1620.1 & 67.85 & 136.3   \\ \hline
ZR 8 & 2387.0 & 1751.7 & 73.39 & 127.1   \\ \hline
ZR 9 & 2387.7 & 1670.7 & 69.95 & 132.5   \\ \hline

 \multicolumn{5}{|c|}{\textbf{ZR tails}} \\ \hline
 
ZR 10 & 2326.2 & 1525.2 & 65.57 & 123.9   \\ \hline
ZR 11 & 2446.8 & 1473.5 & 60.22 & 135.8   \\ \hline
ZR 12 & 2411.3 & 1527.4 & 63.34 & 134.6   \\ \hline

 \multicolumn{5}{|c|}{\textbf{Residual material}} \\ \hline

ZR 13 & 2666.6 & 1501.4 & 56.30 & 147.1   \\ \hline

\end{tabular}
\caption{The loaded mass, the yield, and the exposure of the ZR runs. First the material from the reduction was purified (ZR1 to ZR9), then the ZR tails (ZR10 to ZR12) and finally the recycled tails with the residual Ge (ZR13).}
\label{tab:zrSummary}
\end{table}

A summary of the data from the ZR process is given in Table~\ref{tab:zrSummary}. Here, the run number, the yield, and the exposure are presented for each run. The recycling of the tails continued until there was not enough material to start another ZR. The final ZR yield with continuous recycling, relative to the reduced Ge, was 99.05\%. This includes the last remaining low resistivity tail with a mass of 1153.2~g. The amount lost during cutting of the bars was 87.4~g and the amount lost during etching of the tails was 90.5~g. The etching solution as well as the cutting sludge were collected for later chemical recovery.  

\section{Cosmogenic exposure}
\label{sec:exposure}

Cosmogenic exposure of the enriched Ge has to be minimized at any given time, since the hadronic component of the cosmic radiation produces the radioactive isotopes $^{68}$Ge and $^{60}$Co in Ge, which contributes to the background in the detector spectrum in the region of interest for the $0\upnu\upbeta\upbeta$ decay~\cite{BARABANOV2006115}. Therefore, a significant effort was invested in reducing the cosmogenic exposure of the enriched germanium.

At the beginning of the operation all the enriched germanium dioxide was transported to the shallow-underground laboratory Felsenkeller (Felsenkeller) of the Helmholtz Zentrum Dresden Rossendorf (HZDR) and TU Dresden \cite{bemmerer2015,LUDWIG201924}. The Felsenkeller has an up to 45~m hornblende monazite rock overburden, corresponding to 130~m of water equivalent (m.w.e.). A dedicated area with 75~m.w.e. shielding served as a longer term storage between the processing steps.

The necessary amount of material to start the reduction or ZR processes was regularly transported to IKZ Berlin. At the end of every process the material was transported back to the Felsenkeller, unless it was further processed.

The mean exposure is calculated, since some parts of the germanium underwent more processing cycles than others. Therefore, the exposure in each process step was weighted with the amount of germanium processed (kg$\cdot$h) and at the end the exposure sum was divided by the total mass of the germanium to obtain the mean exposure in hours. Hence, the exposures of the single processes in Tables \ref{tab:red} and \ref{tab:zrSummary} are given in kg$\cdot$h.

\subsection{Exposure in reduction and zone-refining}
To start a reduction process the graphite boats were filled with oxide and the residual powder of the packaging units was stored above ground. Thus, these packages received a higher exposure, which has been taken into account in Table \ref{tab:red}. The exposures of the different processes fluctuated due to technical reasons. After each reduction process the furnace was reloaded with oxide and the Ge bars were either transported to the Felsenkeller or directly used in a ZR process.

\begin{table}[!ht]
\centering
\begin{tabular}{|l|l|l|l|}
\hline
Exposure                    & kg$\cdot$h & h       \\ \hline
Transport Munich - Dresden  & 105.4      & 4.58    \\ \hline
Reduction                   & 1,647.7    & 71.56   \\ \hline
Cutting                     & 130.4      & 5.71    \\ \hline
ZR                          & 1,703.8    & 74.59   \\ \hline
Total                       & 3,587.3    & 156.44  \\ \hline
\end{tabular}
\caption{Summary of the mean cosmogenic exposure contributions weighted by the germanium amount [kg$\cdot$h] and relative to the total mass [h].}
\label{tab:exposure}
\end{table}

ZR was done in parallel to the reduction to minimize material transport. Both the reduction process and a ZR run each lasted about 48~h. After each ZR run the bars were cut, the tails etched, and the material brought back underground.

\subsection{Overall exposure}
The exposure during material processing and transport are summarized in Table~\ref{tab:exposure}. The exposure during the transport from Munich to Dresden was 105.4~kg$\cdot$h. The accumulated weighted exposure during the reduction was 1647.7~kg$\cdot$h which converts to a mean exposure of the 23 kg germanium of 71.56~h. The weighted exposure due to ZR was 1703.8~kg$\cdot$h corresponding to a mean exposure of 74.59~h. Additional exposure of 5.71~h was accumulated when some of the bars had to be brought back for cutting.

At the end, the 23~kg of Ge accumulated a mean exposure of 156.44~h, slightly more than the 126~h achieved with a previous batch for the same operation with an industrial partner \cite{enrBEGe2015}. The difference is mainly due to the larger distance between the processing and storage sites.   
This information will later be used to calculate the exposure of single HPGe detectors as it was done in~\cite{enrBEGe2015}.

\section{Outlook and summary}

For the planned ton-scale $^{76}$Ge 0$\upnu\upbeta\upbeta$ decay experiment LEGEND-1000, large amounts of enriched germanium have to be processed with the highest possible yield, while minimizing the cosmic ray exposure. The stringent requirements and the relatively small batches make it difficult to outsource the task to an industrial company by maintaining full control over the processes. 

To ensure full control over the quality and to maximize the yield, a small scale processing line was setup at IKZ for the reduction of the enriched GeO$_2$. This installed reduction furnace is a complement to the long term effort of IKZ to produce HPGe crystals for fundamental research \cite{abrosimov2020}.

As a first use of the newly built setup the hydrogen reduction of 32~kg enriched GeO$_2$ was performed. Beforehand, we performed a process optimization with special focus on the process efficiency and cosmogenic exposure. The optimization of the reduction process is based on the un-reacted shrinking model, which we used to compute a temperature profile that assures constant reaction rate and the shortest possible process time under the given the constrains. To prevent losses due to the sublimation of the GeO, reduction at temperatures below the recommended 700~\textdegree C was favored. 

We successfully reduced the 32801~g oxide and obtained 23027~g of germanium with an average reduction yield of 99.85\%. While not being processed the material was stored underground in the Felsenkeller laboratory in Dresden. The mean exposure during reduction was 71.56~h. We attempted to recover the smallest amounts of Ge loss in each process step. A total of 107~g of GeO$_2$ powder was recovered which was added to the last reduction process. 

The purity (net charge carrier density) of the Ge bars from the reduction were far from intrinsic (6N), hence ZR had to be performed to reach the desired purity. The mean exposure during ZR was 74.59~h, totaling 156.44~h for the whole processing (including the transport and additional cutting exposure). A summary of the exposure is given in Table \ref{tab:exposure}. 

Major losses that could not be recovered in-house have been collected for later recovery, these include 33.4~g Ge lost during the reduction (potentially in the etching solution), the cutting loss during ZR of 87.4~g, and the etching loss from ZR of 90.5~g.

The total yield after the ZR was 99.05\% and 94.05\% for 6N Ge, which is equivalent to a mass of 21688~g. A summary of the initial and final weights, and yields of the processing steps is presented in Table \ref{tab:summary}.

\begin{table}[!ht]
\centering
\begin{tabular}{|l|l|l|}
\hline
                     & Weight [g] & Yield [\%] \\ \hline
GeO$_2$ measured     & 32801.2    & -          \\ \hline
Ge expected          & 23059.9    & -          \\ \hline
After reduction      & 23026.5    & 99.85      \\ \hline
After ZR, total      & 22841.8    & 99.05      \\ \hline
Thereof 6N           & 21688.6    & 94.05      \\ \hline
Tail < 50 $\Omega$cm & 1153.2     & 5.00       \\ \hline
\end{tabular}
\caption{Summary of the initial and final weights, and yields of the reduction and ZR.}
\label{tab:summary}
\end{table}

Finally, we developed a process for the hydrogen reduction of germanium dioxide enriched in $^{76}$Ge as part of our contribution to the LEGEND experiment. We were able to fulfill the yield and exposure requirements with a laboratory-scale setup processing about a kilogram Ge at once. 
The hardware setup and the process optimization have been successfully completed. In the future, the described setup can be transferred to an underground facility if further reduction of the cosmogenic exposure is needed. The most challenging associated tasks include the secure supply and control of gases, especially hydrogen, and the handling of chemicals.

With an overall Ge yield above 99\% we succeeded in minimizing the loss of Ge during processing. The losses will be further reduced when the Ge in the cutting sludge and the acid solutions will be recycled. The apparently small gain will be a major contribution when it will be applied to the ton-scale LEGEND-1000 experiment.

\acknowledgments

We acknowledge the support of the German Federal Ministry for Education and Research (BMBF) within the collaborative project under the grant number 05A2017 (GERDA/LEGEND). These results are part of the GemX project that has received funding from the European Research Council (ERC) under the European Union’s  Horizon 2020 research and innovation programme  (Grant agreement No. 786430). We also gratefully acknowledge the support of D. Bemmerer and the Helmholtz Zentrum Dresden Rossendorf for the possibility to store the germanium material in the Felsenkeller underground laboratory, and S. Nisi and F. Ferella from the Laboratori Nazionali del Gran Sasso (LNGS) for the elemental analysis of the enriched germanium oxide.

\bibliographystyle{unsrt}
\bibliography{main}

\end{document}